# Relativity and Diversity of Strong Coupling to Measured Subsystems

*Renming Liu,[†] Yi-Cong Yu,[‡] and Xue-Hua Wang[†,*]*

[†]State Key Laboratory of Optoelectronic Materials and Technologies, School of Physics, Sun Yat-sen University, Guangzhou 510275, China

[‡]School of Physics and Optoelectronic Engineering, Foshan University, Foshan 528000, China

**ABSTRACT:** The strong coupling between two subsystems consisting of quantum emitters and photonic modes, at which the level splitting of mixed quantum states occurs, has been a central subject of quantum physics and nanophotonics due to various important applications. The spectral Rabi-splitting of photon emission or absorption has been adopted to experimentally characterize the strong coupling under the equality assumption that it is identical to the level splitting. Here, we for the first time reveal that the equality assumption is not valid. It is the invalidity that results in the relativity and diversity of the strong coupling characterized by the spectral Rabi-splitting to the measured subsystems, highly correlated with their dissipative decays. The strong coupling is easier to be observed from the subsystem with larger decay, and can be classified into pseudo-, dark-, middle-, and super-strong interaction regimes. We also suggest a prototype of coupled plasmon-exciton system for possibly future experiment observations on these novel predictions. Our work brings new fundamental insight to the light-matter interaction in nanostructures, which will stimulate further researches in this field.

**KEYWORDS:** Plasmons, Quantum emitter, Strong coupling, Relativity, Spectral Rabi-splitting, Dissipation.

The strong coupling between a quantum emitter (QE) and the electromagnetic mode is a distinct regime of light−matter interactions, which has been one of the central subjects of quantum physics and nanophotonics. In such a strong interaction regime, energy exchange between the two subsystems is on time scales faster than their respective dissipations, manifests itself as mode hybridization and level splitting of the mixed quantum states that are part light and part matter. This not only allows for the testbed study of fundamental scientific problems in cavity quantum electrodynamics (CQED), such as quantum entanglement,[1,2] Bose−Einstein condensates[3−5] and superfluidity,[6] but also holds great potential for important applications including quantum computing,[7-9] quantum information storage and processing,[10,11] as well as polariton lasing.[12]

The strong coupling, a fundamental quantum effect, intrinsically reflects the internal level splitting of the quantum state in the coupled subsystems. Experimentally, it has been characterized by the spectral Rabi-splitting in either emission, or absorption/scattering spectrum from the observed subsystem based on the equality assumption of the spectral Rabi-splitting and the level splitting. Traditionally, this quantum effect is demonstrated by measuring the vacuum Rabi splitting of the emission spectra from the QE in the CQED systems including various optical dielectric microcavities at cryogenic temperatures in ultrahigh vacuums.[13−19] Recently, various novel plasmonic nanostructures have been increasingly used to realize the room-temperature strong coupling beyond the optical dielectric microcavities.[20−26] In particular, by using the localized surface plasmon resonance with deep subwavelength mode confinement, recent experiments have demonstrated the room-temperature strong coupling even down to the single QE level.[27–29] Different from the traditional CQED systems, the room-temperature strong coupling between the plasmon modes and the QEs is usually characterized by the spectral Rabi-splitting of the photon absorption/scattering from the plasmon modes. However, it is controversial whether the spectral Rabi-splitting from plasmon mode truly reflects the strong coupling of the coupled plasmon-QE systems.[26,30,31] Thus, the fundamental scientific issue about the aforementioned equality assumption remains to be answered. Moreover, due to the lack of investigations on the absorption/scattering spectra from the QEs, it is also unclear whether the strong coupling characterized by the spectral Rabi-splitting depends on the probed subsystem.

In this work, we will firstly derive the analytical formulae and critical criteria of both the level splitting and the spectral Rabi-splitting of the coupled plasmon-QE system based upon the quantum theory, and prove that the quantum levels of the coupled plasmon-QE system are identical to the singular-point energies of the complex absorption-spectrum function, rather than the spectral Rabi-splitting energies. Then, by comparing these critical criteria, we reveal that the relativity and diversity of the strong coupling characterized by the spectral Rabi-splitting of the photon absorption to the measured subsystems are highly correlated with their dissipative decays. We also present the identification method of the spectral Rabi-splitting signal channel and correct measurement method of g-factor for future experiment researches. Finally, we suggest a prototype of coupled plasmon-exciton system and demonstrate it a good candidate for possibly future experiment observations on the relativity and diversity by FDTD numerical simulations.

**Level splitting and optical absorption in strongly coupled plasmon-QE system.** It is well known that the coupling described by the JC Hamiltonian is inherent to all interacting photon-QE systems. However, the Rabi splitting can be blurred out by spontaneous decay and dephasing in the system. By phenomenologically introducing the losses associated with incoherent processes, the dressed eigenlevels of the coupled systems can be obtained by diagonalizing the following non-Hermitian Hamiltonian[13]

$$H_{loss} = \begin{pmatrix} \varepsilon_d - i\gamma_d & g_{dc} \\ g_{dc} & \varepsilon_c - i\gamma_c \end{pmatrix} , \qquad (1)$$

where $\gamma_d$ and $\gamma_c$ are half decay linewidths of the plasmon mode and the QE, respectively; $g_{dc}$ (called as g-factor) is the coupling energy between them. Eq 1 can be generalized to the two bright states of $N$ identical QEs coupled to the plasmon mode in the single excitation case. In fact, such an ensemble at weak excitations behaves as a giant harmonic oscillator giving rise to the coupling energy of $G = \sqrt{N} g_{dc}$.[32,33] By diagonalizing of this Hamiltonian, the energies of the two mixed levels can be obtained as[13,19,34]

$$\varepsilon_\pm^l = (\varepsilon_c + \varepsilon_d)/2 - i(\gamma_d + \gamma_c)/2 \pm \Delta_{ls}, \ \Delta_{ls} = \sqrt{g_{dc}^2 + [\delta + i(\gamma_d - \gamma_c)]^2/4}, \qquad (2)$$

where $\delta = \varepsilon_d - \varepsilon_c$ is the detuning. At resonance ($\delta = 0$), the condition of the level splitting is

$$g_{dc}^2 > g_{ls}^2 = (\gamma_d - \gamma_c)^2/4, \qquad (3)$$

under this condition, the coupling between two subsystems is strong enough to generate two mixed state levels. Up to now, the strong coupling has been characterized by the spectral Rabi-splitting,[33,35,36] which implies the equality assumption of the spectral Rabi-splitting and the level splitting.

In order to test this equality assumption, we now investigate the spectral Rabi-splitting in the optical absorption/scattering from the coupled plasmon-QE systems. To begin with, we describe the plasmon mode with bosonic annihilation and creation operators and the QE as a fermion with only two possible states (ground and excited). The Hamiltonian describing the coupled quantum system can be written as[37]

The Hamiltonian describing the noninteracting evolution of the fermion and the plasmon is $H_0 = \varepsilon_d d^+ d + \varepsilon_c c^+ c$, where $d$ and $c$ ($d^+$ and $c^+$) are the annihilation (creation) operators for the plasmon mode and the quantum emitter fermion with energies $\varepsilon_d = \hbar\omega_d$ and $\varepsilon_c = \hbar\omega_c$. The plasmon-QE interaction is modeled by the Hamiltonian $H_{int} = -g_{dc}\left[d^+ c + c^+ d\right]$. To describe the finite lifetimes of the plasmon and fermion, we add a term to the Hamiltonian that contains the inelastic interactions with a continuum of modes, $H_{decay}$ (see Section S1.1 in SI). Under the condition of weak light excitation, the photon absorption spectrum observed from the plasmon mode ($\sigma_d(\omega)$) and the QE ($\sigma_c(\omega)$) can be deduced from the Zubarev's Green function $\langle\langle d; d^+ \rangle\rangle_{\omega+i0^+}$ and $\langle\langle c; c^+ \rangle\rangle_{\omega+i0^+}$ as

$$\sigma_d(\omega) \propto -\text{Im}\frac{\hbar\omega - \varepsilon_c + i\gamma_c}{(\hbar\omega - \varepsilon_d + i\gamma_d)(\hbar\omega - \varepsilon_c + i\gamma_c) - g_{dc}^2}, \tag{5}$$

and

$$\sigma_c(\omega) \propto -\text{Im}\frac{\hbar\omega - \varepsilon_d + i\gamma_d}{(\hbar\omega - \varepsilon_d + i\gamma_d)(\hbar\omega - \varepsilon_c + i\gamma_c) - g_{dc}^2}, \tag{6}$$

respectively. For the coupled plasmon-QE system, the observed photon absorption spectrum can be described as

$$\sigma(\omega) = A_1 \cdot \sigma_d(\omega) + A_2 \cdot \sigma_c(\omega), \tag{7}$$

where $A_1$ and $A_2$ ($A_1+A_2=1$) are the weight factors observed from the two channels of the plasmon mode and QE, respectively. At resonance ($\omega_d=\omega_c$), letting $\frac{d\sigma(\omega)}{d\omega}=0$, we obtain the criterion for the spectral Rabi-splitting observed from the plasmon mode channel (Ch1) (i.e. $A_1=1$, $A_2=0$) as

$$g_{dc}^2 > g_{pl}^2 = \frac{\gamma_c^2}{2(1+\gamma_d/2\gamma_c)}, \qquad (8)$$

and the corresponding energies as

$$\varepsilon_\pm^p = \varepsilon_c \pm \Delta_{pl}, \ \Delta_{pl} = \sqrt{g_{dc}(1+\gamma_c/\gamma_d)\cdot(g_{dc}^2+\gamma_c\gamma_d)^{1/2}-(g_{dc}^2+\gamma_c\gamma_d)\cdot\gamma_c/\gamma_d}. \qquad (9)$$

Similarly, by letting $\frac{d\sigma(\omega)}{d\omega}=0$ at $A_1=0$ and $A_2=1$, the criterion for the spectral Rabi-splitting observed from the QE channel (Ch2) can be given as

$$g_{dc}^2 > g_{qe}^2 = \frac{\gamma_d^2}{2(1+\gamma_c/2\gamma_d)}, \qquad (10)$$

and the corresponding energies as

$$\varepsilon_\pm^q = \varepsilon_d \pm \Delta_{qe}, \ \Delta_{qe} = \sqrt{g_{dc}(1+\gamma_d/\gamma_c)\cdot(g_{dc}^2+\gamma_c\gamma_d)^{1/2}-(g_{dc}^2+\gamma_c\gamma_d)\cdot\gamma_d/\gamma_c}. \qquad (11)$$

**Relativity and diversity of strong coupling to the measured subsystems.** We now turn to discuss the relativity and diversity of the strong coupling characterized by the spectral Rabi-splitting. Without loss of generality, we consider the situation of $\gamma_d > \gamma_c$ (i.e., the decay of the plasmon mode is larger than that of the QE) in the following discussions. Because $g_{qe} > g_{pl}$ is always fullfiled according to eqs 8 and 10 under the condition of $\gamma_d > \gamma_c$, this implies that it is easier for the subsystem with larger dissipation to observe the strong coupling represented by the spectral Rabi-splitting than for the subsystem with smaller loss, which is in vast contrast to the intuition. The physics mechanism of this phenomenon can be understood as follows. The strong coupling between the two subsystems requires a complete cycle of energy exchange, as shown in Figure 1a, that depicts a general model of a plasmonnano-subsystem (PNS) interacting with a QE-subsystem: one subsystem is driven firstly by external probing, it transfers an energy quantum to the second subsystem and excites it by the coupling between them, and then the excited second subsystem will transfer the energy quantum back to the first one. However, the energy transfers between the two subsystems and the dissipation via the subsystems are two competitive processes. A complete cycle of the energy exchange will be interrupted if the decay rate of the second subsystem is greater than the energy transfer rate. Therefore, the smaller the dissipation of the second subsystem is, the easier it is to realize a complete cycle of the energy transfer, and the subsystem with larger decay is a good channel for probing the strong coupling.

We now reveal the diversity of the strong coupling by carefully considering the two cases of $\gamma_c < \gamma_d/2$ and $\gamma_c > \gamma_d/2$. For $\gamma_c < \gamma_d/2$, the relationship of $g_{pl} < g_{ls} < g_{qe}$ is satisfied. In this case, the strong coupling can significantly be divided as the following three regions: (i) the pseudo-strong coupling region of $g_{pl} \leq g_{dc} < g_{ls}$, where the spectral Rabi-splitting can be observed in the absorption spectrum of the plasmon mode (labeled as Ch1 in Figure 1a) while no level splitting of the mixed quantum states occurs; (ii) the middle-strong coupling regime of $g_{pl} < g_{ls} \leq g_{dc} < g_{qe}$, in which the level splitting of the mixed quantum states appears while the spectral Rabi-splitting can be observed only in absorption spectrum of the plasmon mode (Ch1), and (iii) the super-strong coupling regime of $g_{pl} < g_{ls} \leq g_{qe} < g_{dc}$, where not only the level splitting of the mixed quantum states appears, but also the spectral Rabi-splitting can be observed in absorption spectrum of both the plasmon mode (Ch1) and the QE (labeled as Ch2 in Figure 1a).

For $\gamma_d/2 < \gamma_c < \gamma_d$, the relationship of $g_{ls} < g_{pl} < g_{qe}$ is fullfiled. Similarly, three different strong coupling regions can be found: (i) the dark-strong coupling region of $g_{ls} < g_{dc} < g_{pl} \leq g_{qe}$, where level splitting of the mixed quantum states occurs while the spectral Rabi-splitting cannot be observed in the absorption spectrum from the Ch1 or Ch2; (ii) the middle-strong coupling region of $g_{ls} < g_{pl} < g_{dc} \leq g_{qe}$, where the level splitting occurs and the spectral Rabi-splitting can be observed in the absorption spectrum only from the Ch1; and (iii) the super-strong coupling region of $g_{ls} < g_{pl} < g_{qe} < g_{dc}$, where not only the level splitting appears, but also the spectral Rabi-splitting can be observed in absorption spectra from both the Ch1 and Ch2.

Next, we give the numerically calculated results to demonstrate the strong-coupling relativity and diversity with the typical values of $\gamma_d = 75$ meV and $\gamma_c = 25$ meV respectively for the plasmon modes and excitons at room temperature,[21−25] corresponding to $\gamma_c < \gamma_d / 2$, $g_{pl} = 11.18$ meV, $g_{ls} = 25.00$ meV, and $g_{qe} = 49.10$ meV. From Figure 1b, we clearly observe the different strong coupling regimes for the different coupling energies $g_{dc}$: (I) the pseudo-strong coupling regime when $g_{pl} < g_{dc} = 22$ meV $< g_{ls}$, in which the spectral splitting can be observed from the Ch1 although no level splitting of the quantum state occurs; (II) the middle-strong coupling regime when $g_{ls} < g_{dc} = 45$ meV $< g_{qe}$, in which the level splitting of the quantum state appears and the spectral splitting width from the Ch1 increases, but there is still no the spectral Rabi-splitting from the Ch2; (III) the super-strong coupling regime when $g_{dc} = 70$ meV $> g_{qe} > g_{ls}$, in which the spectral Rabi-splitting in absorption spectra from both the Ch1 and Ch2 can be clearly observed. Nevertheless, the Rabi splitting width observed from Ch1 is much larger than that observed from the Ch2, demonstrating a relativity of strong coupling to the measured subsystem. It should be mentioned that the above situations will be reversed when $\gamma_c$ and $\gamma_d$ exchanged their values (see Figure S1 in SI), indicating that in a coupled plasmon-QE system with $\gamma_d < \gamma_c$, spectral Rabi-splitting in the absorption will be more easily observed from Ch2 rather than from Ch1. Along with great efforts have been made to reduce the losses of plasmon modes,[38-45] plasmonic nanostructures with low dissipations will become increasingly available in the near future. Therefore, it is important to reveal the optical response properties of the QE strongly coupled to the plasmon modes with low dissipation.

From above discussions, the relativity and diversity of the strong coupling are highly dependent on the relative dissipation of the measured subsystems. Figure 1c gives the observed spectral Rabi-splitting and the level splitting of the quantum state as functions of $\gamma_c/\gamma_d$. At the point of $\gamma_c/\gamma_d = 1$, the spectral Rabi-splitting observed in absorption spectra from the two subsystems are always the same and smaller than the level splitting of the quantum states. At the points of $\gamma_c/\gamma_d = 0.80$ and 1.25, however, the spectral Rabi-splittings in the absorption respectively from Ch1 and Ch2 are equal to the level splitting. Aside from these two special points, the spectral Rabi-splittings are quite different from the level splitting. For example, in the condition of $0.8 < \gamma_c/\gamma_d < 1$, the Rabi splitting is more easily observed in absorption from the Ch1, and is smaller than the level splitting. On the contrary, in the region of $\gamma_c/\gamma_d < 0.8$ the spectral Rabi-splitting from this channel is larger than the level splitting; especially in the region of $\gamma_c/\gamma_d < 0.5$, the spectral Rabi-splitting is more than 10% larger than the level splitting. Due to the commutative symmetry of $\gamma_d$ and $\gamma_c$ in above calculations, similar situations can be seen in the region of $\gamma_c/\gamma_d > 1$, and in this case the spectral Rabi-splitting is more easily observed in the absorption spectrum from the Ch2.

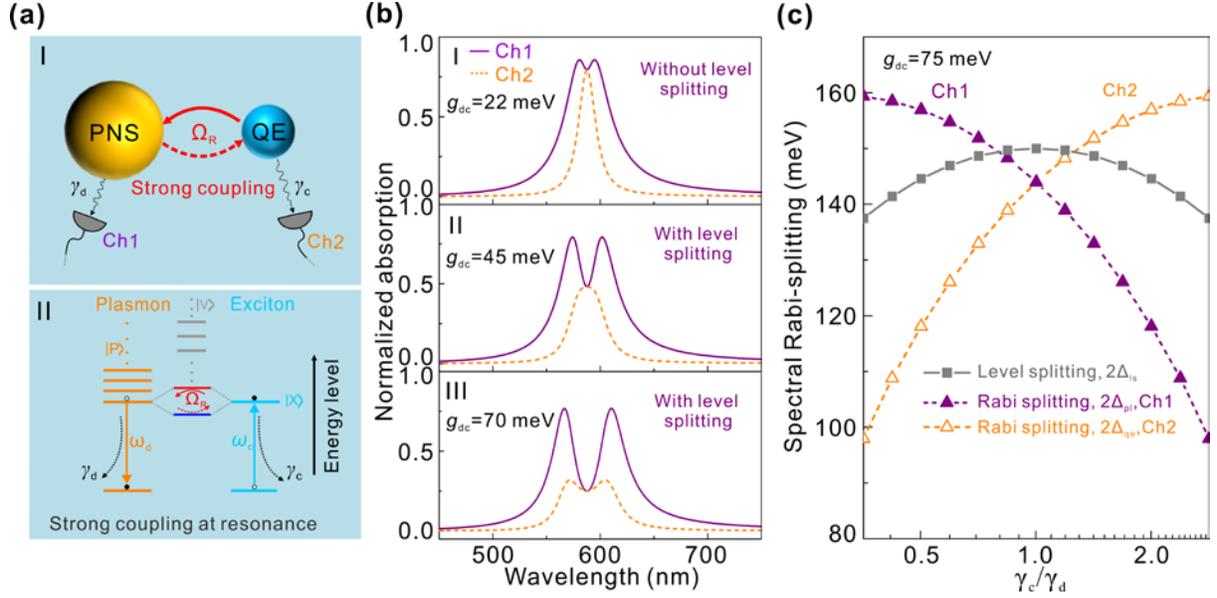

**Figure 1.** The relativity and diversity of strong coupling to the measured subsystems in the coupled plasmon-QE system. (a) (I) Description of a QE strongly coupled to a PNS. (II) Energy diagrams exemplifying the plasmon–QE strong coupling system with the Rabi frequency, $\Omega_R$; $|X\rangle$ represents a single exciton state; the plasmon system is represented by a plasmonic mode as $|P\rangle$; and the continuum of vacuum states is denoted as $|V\rangle$. (b) Calculated absorption spectra of the coupled plasmon-QE system respectively observed from Ch1 (solid purple lines) and Ch2 (dashed orange lines) for three different $g_{dc}$-factors. In calculations, $\varepsilon_c = \varepsilon_d = 2.11$ eV. (c) Spectral Rabi-splitting observed in the absorption spectra, respectively from Ch1 and Ch2 of the strongly coupled plasmon-QE system at different $\gamma_c/\gamma_d$. In calculations, $g_{dc} = 75$ meV, $\gamma_c + \gamma_d = 120$ meV.

**Identification of spectral Rabi-splitting signal channel and correct measurement of $g$-factor.** It is necessary for future experiment investigation on the predicted relativity and diversity to find the identification method of signal channel that can discriminate which channel (i.e. subsystem) the spectral Rabi-splitting signal comes from. Fortunately, the numerators in the formulae eqs 5 and 6 of the absorption spectra provide us such a method. If the spectral Rabi-splitting signal comes from the plasmonic mode, its dip should nearly keep unchanged in the vicinity of the exciton transition $\omega_c$, and does not vary with the detuning of the plasmon mode frequency. Conversely, if the spectral Rabi-splitting signal comes from the QEs, its dip should nearly be in the vicinity of the plasmon mode frequency $\omega_d$, and shift with the detuning of the plasmon mode frequency.

The measurement of the *g*-factor is very important for judging the strong coupling. The existing measurement method of the *g*-factor is that, firstly, the spectral Rabi-splitting is experimentally obtained from the emission/absorption spectrum, and then the *g*-factor is calculated by inserting the spectral Rabi-splitting into the level splitting eq 2.[13,19,34] This method is based on the equality assumption of the spectral Rabi-splitting and the level splitting. However, we have revealed that the two new hybrid levels (eq 2) are exactly identical to the singular-point energies of the complex absorption-spectrum functions (i.e., corresponding to the zero point of the denominators) shown in eqs 5 and 6, and evidently differ from the spectral Rabi-splitting energies obtained by eq 9 or 11. This implies that the equality assumption is invalid. Therefore, the measurement of the *g*-factor should build on eq 9 or 11 depending on which channel (i.e. subsystem) the spectral Rabi-splitting signal comes from, instead of eq 2.

**Proposed plasmonic grating structure for demonstrating the strong-coupling relativity.** For guiding the experimental demonstration of the strong-coupling relativity to the different channels (subsystems), we propose a coupled plasmon-exciton system made of an excitonic thin-layer and a plasmonic grating structure (PGS).[38] The PGS consists of a dielectric grating on a gold nanofilm deposited on a silicon substrate (Figure 2a). Figure 2b shows the absorption spectra of the PGS under normal and oblique incidence for TM polarization. Under normal incidence (α=0°), only one absorption peak appears at ~701 nm that corresponds to the excitation of the localized surface plasmon (LSP) mode P0. Under the oblique incidence of $α = 17.5°$, however, two surface plasmon polariton (SPP) modes with the absorption peaks at ~597 (P1) and ~833 nm (P2) emerge, in contrast to the absorption spectra of the bare dielectric grating and the Au nanofilm on the silicon substrate (dashed green and black curves in Figure 2b, respectively). Compared to the linewidth ~80 meV of the SPP mode P1 at ~597 nm, the SPP mode P2 at ~833 nm has a much narrower linewidth of ~10 meV, and which can be further compressed by increasing $α$. In addition, the resonance wavelength of the SPP mode P2 can be linearly tuned by changing α (Figure 2c (I)). Different from the SPP modes, the resonance wavelength of the LSP mode P0 is tuned by changing the thickness of the grating under normal incidence. Figure 2c (II)) shows that as the thickness of the grating increases from 150 to 350 nm, the resonance wavelength of the LSP mode P0 is red-shifted from ~730 to 760 nm. Figure 2d depicts the electric field enhancement ($|E/E_0|$) inside and around the grating media at the plasmon resonance wavelengths of 701 (P0), 597 (P1) and 833 nm (P2), respectively. Contrary to the EF distribution of the LSP mode P0 that is highly localized in and around the grating (Figure 2d (I)), the EFs of the SPP modes P1 and P2 are mainly distributed around the dielectric grating, showing enhanced propagation properties. In particular, the SPP mode P2 has a much better propagation property than that of SPP mode P1.

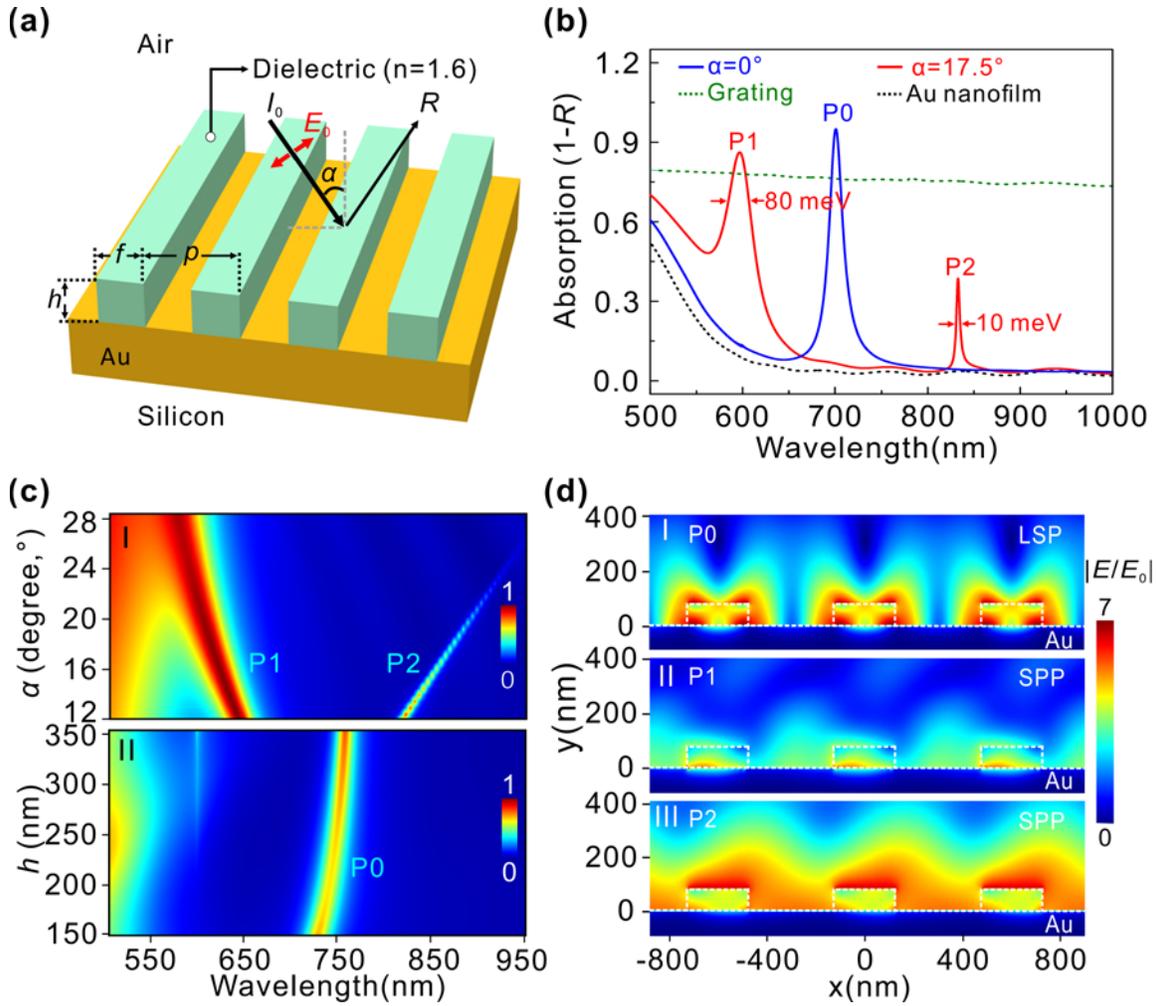

**Figure 2.** Optical response of the plasmonic grating structure (PGS). (a) Schematic of the PGS consisting of a dielectric grating on a gold film (100-nm thick) deposited on a silicon substrate. (b) FDTD calculated absorption spectra (1-*R*, *R*: reflection) of the PGS for dielectric grating with the period of *p*=600 nm, the thickness of *h*=80 nm, and the fill factor of *f* = 0.4. The solid blue and red curves are the absorption spectra of the plasmonic structure excited by TM polarized incident light at *α*=0° and 17.5°, respectively. The dashed green and black curves are the absorption spectra of the dielectric grating and Au nanofilm on the bulk Si substrate excited by TM polarized incident light at *α*=17.5°, respectively. (c) Calculated absorption spectra of the PGS: (I) at *h*=80 nm for varying *α*, and (II) at a normal incidence (*α*=0°) for different *h*. (d) The EF enhancement ($|E/E_0|$) for the LSP mode P0 at (I) $\alpha = 0°$ and for the SPP modes (II) P1 (597 nm) and (III) P2 (833 nm) at $\alpha = 17.5°$.

**Classical simulation of strong coupling with $\gamma_d > \gamma_c$ and its quantum identification.** In order to facilitate the realization of strong coupling between the LSP mode P0 and the exciton transition, we consider that an excitonic thin-layer (dye molecules or two-dimensional transition-metal dichalcogenides) with a transition frequency of $\omega_0 = 1.642$ eV and $2\gamma_c = 35$ meV is embedded between the dielectric grating and gold nanofilm (Figure 3a). Generally, the optical properties of such hybrid plasmon-exciton systems can be simulated using the finite-difference time-domain (FDTD) or full wave finite element calculations, in which the excitonic thin-layer is modeled as the dielectric layer with the classical Lorentz permittivity[24-26,28] (see eq S48 in Section S2 of SI). In the following calculations, we consider the excitonic thin-layer of 5-nm thickness. Apparently, the PGS and the excitonic thin-layer are each other's dielectric background, which leads to the slight shifts of both the frequency $\omega_d$ of the LSP mode P0 and the resonant transition frequency $\omega_c$ (i.e. absorption frequency) of the excitonic thin-layer (Figure 3b and Supplementary Fig. S2). Figure S2b displays their variations with the grating thickness $h$. The FDTD simulations show that the resonant coupling between the two subsystems appears at $h=234$ nm corresponding to $\omega_d = \omega_c = 1.660$ eV, at which the spectral Rabi-splitting is displayed by the solid red curve in Figure 3c (I). However, these classical simulations cannot discriminate which channel (or subsystem) the spectral Rabi-splitting signal comes from. In order to clearly identify it, we try to reproduce the absorption spectra by eq 7 based upon quantum theory. To do this, we have to obtain the coupling energy $g_{dc}$ and the linewidth $2\gamma_d$ of the LSP mode P0 besides the parameters $\omega_d = \omega_c = 1.660$ eV, $2\gamma_c = 35$ meV. $g_{dc}$ can be calculated by eq 9 or 11 according to the spectral Rabi-splitting $2\Delta = 50$ meV shown by the solid curve in Figure 3c (I). Combining eqs 8 and 10 with the condition of $\gamma_d > \gamma_c$, we deduce the spectral Rabi-splitting signal originates from the LSP mode P0. Therefore, we obtain $g_{dc} = 24.63$ meV by uing eq 9 with $2\Delta_{pl} = 50$ meV. The influence of the excitonic thin-layer on the LSP mode P0 can approximately be simulated by replacing the Lorentz permittivity $\varepsilon_{ex}(\omega)$ with its high-frequency component $\varepsilon_{ex}(\infty) = 2.6$ (see eq S48), as shown in Figure S2b. The the absorption spectrum of the LSP mode P0 obtained by this approximation at $h=234$ nm is depicted in Figure 3c(II) with the linewidth of $2\gamma_d \sim 51$ meV and the resonant absorption frequency of $\omega_d' = 1.655$ eV that is very close to $\omega_d = \omega_c = 1.660$ eV. By inserting $g_{dc} = 24.63$ meV, $\omega_d = \omega_c = 1.660$ eV, $\gamma_d = 25.5$ meV and $\gamma_c = 17.5$ meV into eq 7 with setting $A1=1$ and $A2=0$, the quantum absorption spectrum is depicted by the dash blue curve in Figure 3c(I), which very well reproduces the classical FDTD results.

On the other hand, eq 5 shows that if the absorption signal comes from the plasmonic mode, the dip of the spectral Rabi-splitting should always be near the exciton frequency $\omega_c$, and does not vary with the tuning of the plasmon mode. As the grating thickness, $h$, is varied from 60 to 300 nm, the detuning ($\delta = \varepsilon_d - \varepsilon_c$) between the LSP mode P0 and the exciton transition can be well controlled (Figure 3b). The FDTD-simulated absorption spectra of the coupled system at $h = 196$, 234 and 256 nm are shown in Figure 3d, and they are very well reproduced by quantum theory eq 7 with $A1=1$ and $A2=0$ (*i.e.* eq 5) for the corresponding LSP mode frequencies at 1.670, 1.660 and 1.648 eV, respectively (Figure 3e). Indeed, the dips of spectral Rabi-splitting nearly keep unchanged in the vicinity of the exciton transition $\omega_c$, which further confirms that the spectral Rabi-splitting signal comes from the LSP mode P0.

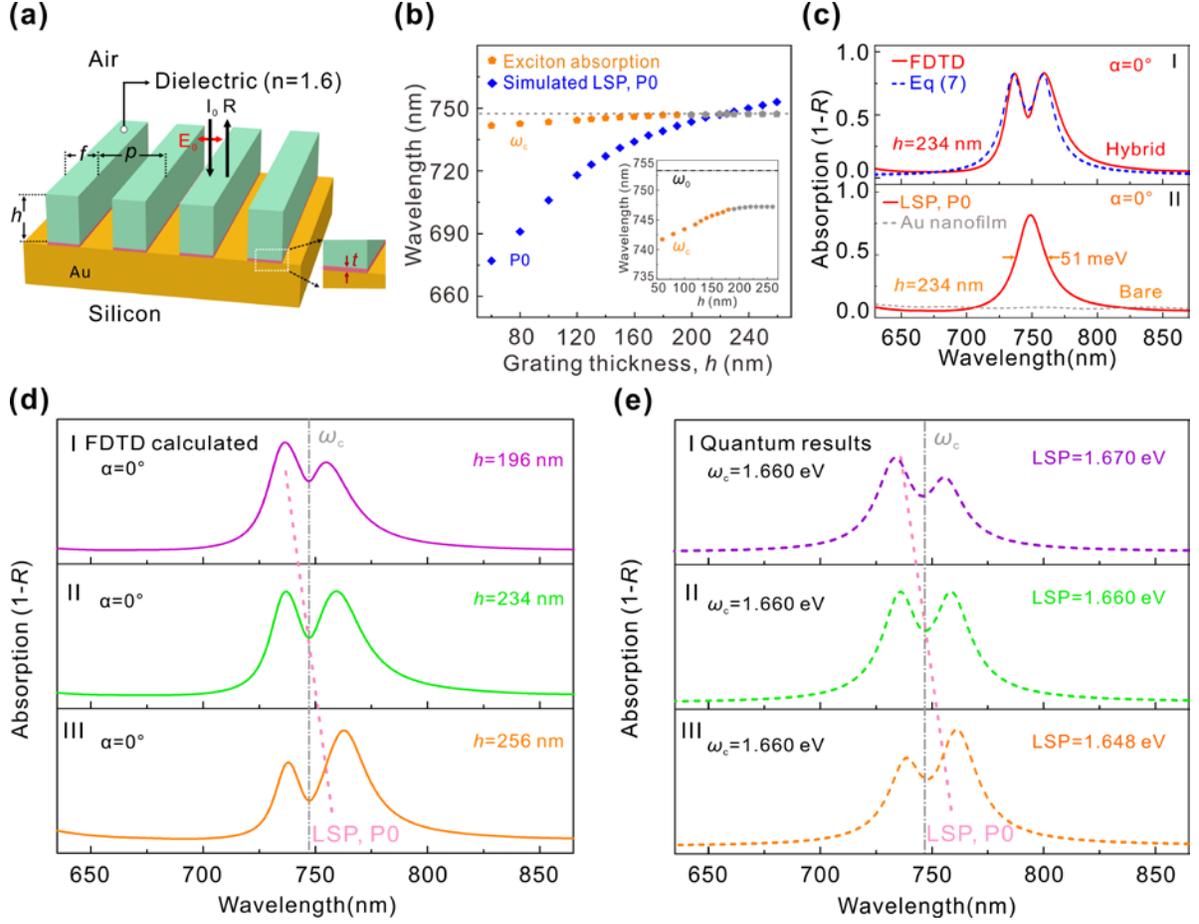

**Figure 3.** Strong coupling between the LSP mode P0 and exciton. (a) Schematic of the hybrid plasmonic structure formed by introducing an excitonic thin-layer ($t = 5$ nm) between the Au nanofilm and the dielectric grating of the PGS. (b) The excitonic transition, $\omega_c$, and the resonance wavelength of the LSP mode P0 as functions of $h$. The gray pentagons are $\omega_c$ measured from the dips of the spectral Rabi-splitting (Figure S2a in SI). The inset is a more intuitive demonstration of the change for $\omega_c$ versus $h$. (c) Absorption spectra of (I) the hybrid plasmonic structure including the excitonic thin-layer and (II) the bare PGS with $h$=234 nm, calculated by FDTD for TM polarized incident light at $\alpha$=0°. The dashed black curve in (I) is the normalized absorption spectrum of the hybrid plasmonic nanostructure calculated using eq 7 with $A1$=1 and $A2$= 0, which is normalized to the maximum absorption calculated at $g_{dc}$=0. (d)

FDTD calculated absorption spectra for the hybrid plasmonic structure with $h$=196, 234 and 256 nm, respectively. (e) Normalized absorption spectra calculated using eq 7 with $A1$= 1 and $A2$= 0 for the hybrid nanostructure with LSP mode frequencies corresponding to the parameters shown in (d). In the calculations, $g_{dc}$ =24.63 meV is obtained from extracted from eq 9 with $\Delta_{pl}$ =25 meV.

**Classical simulation of strong coupling with $\gamma_d < \gamma_c$ and its quantum identification.** To further verify the relativity of the strong coupling, the excitonic thin-layer with a transition frequency of $\omega_0 = 1.510$ eV and $2\gamma_c = 35$ meV as reset and placed on the grating of the PGS to interact with the SPP mode P2 (Figure 4a), which has a much narrower linewidth compared to the exciton transition. Similarly, when the excitonic thin-layer was put on the surface of the dielectric grating, slight shifts of both the frequency $\omega_d$ of the SSP mode and the transition frequency $\omega_c$ of the excitonic thin-layer in the hybrid plasmonic structure can also be observed as above mentioned (Figure 4b and Supplementary Figure S3). By precisely controlling the α, the resonance detuning between the SPP mode and the excitonic transition can be achieved. The simulated results show that when the α was set at 18.2°, corresponding to $\omega_d = \omega_c = 1.485$ eV, the resonant coupling between the two subsystems appears and the spectral Rabi-splitting is displayed by the solid red curvein Figure 4c(I). To discriminate which channel the spectral Rabi-splitting of photon absorption comes from, we also try to reproduce this absorption spectra by using eq 7. To do this, we need to obtain the coupling energy $g_{dc}$ and the linewidth ($2\gamma_d$) of the bare SSP mode P2 besides the parameters $\omega_d = \omega_c = 1.485$ eV and $2\gamma_c = 35$ meV. By considering the eqs 8 and 10 with the condition of $\gamma_c > \gamma_d$, we predict that the spectral Rabi-splitting signal of this coupled system comes from the excitonic channel (Ch2), other than the SSP mode P2 (Ch1).

Then, we can obtain $g_{dc}$=18.25 meV by eq 11 according to the spectral Rabi-splitting $2\Delta_{qe}$=38 meV shown in Figure 4c(I). The linewidth of the bare SSP mode P2 at $\alpha$=18.2° is obtained as $2\gamma_d$ ~10 meV from Figure 4c(II). By inserting $g_{dc}$=18.25 meV, $\omega_d = \omega_c$ =1.485 eV, $\gamma_d$=5 meV and $\gamma_c$= 17.5 meV into eq 7 with setting $A_1$=0 and $A_2$=1, the quantum absorption spectrum is shown by the dash curve in Figure 4c(I), which well reproduces the classical FDTD results.

In order to further confirm that the spectral Rabi-splitting signal of this coupled system originates from the excitonic channel Ch2 (i.e. QEs), we examine the variation of the Rabi-splitting dip with the detuning frequency of the SPP mode P2. The FDTD-simulated absorption spectra of the coupled system at $a$ =17, 18.2 and 19° are depicted in Figure 4d, corresponding to SSP resonance frequencies of 1.500, 1.485 and 1.475 eV, respectively. Once again, they are well reproduced by quantum eq 7 with setting $A_1$=0 and $A_2$=1 at the corresponding SSP resonance frequencies, as shown in Figure 4e. It is interesting to notice that the dips of the spectral Rabi-splitting always keep around the plasmon resonance frequency $\omega_d$ and shift with $\omega_d$. According to the abovementioned identification method of the spectral Rabi-splitting signal channel, we can conclude that the strong coupling signal in Figure 4c(I) indeed comes from the QEs' absorption, rather than the absorption of the SPP mode P2.

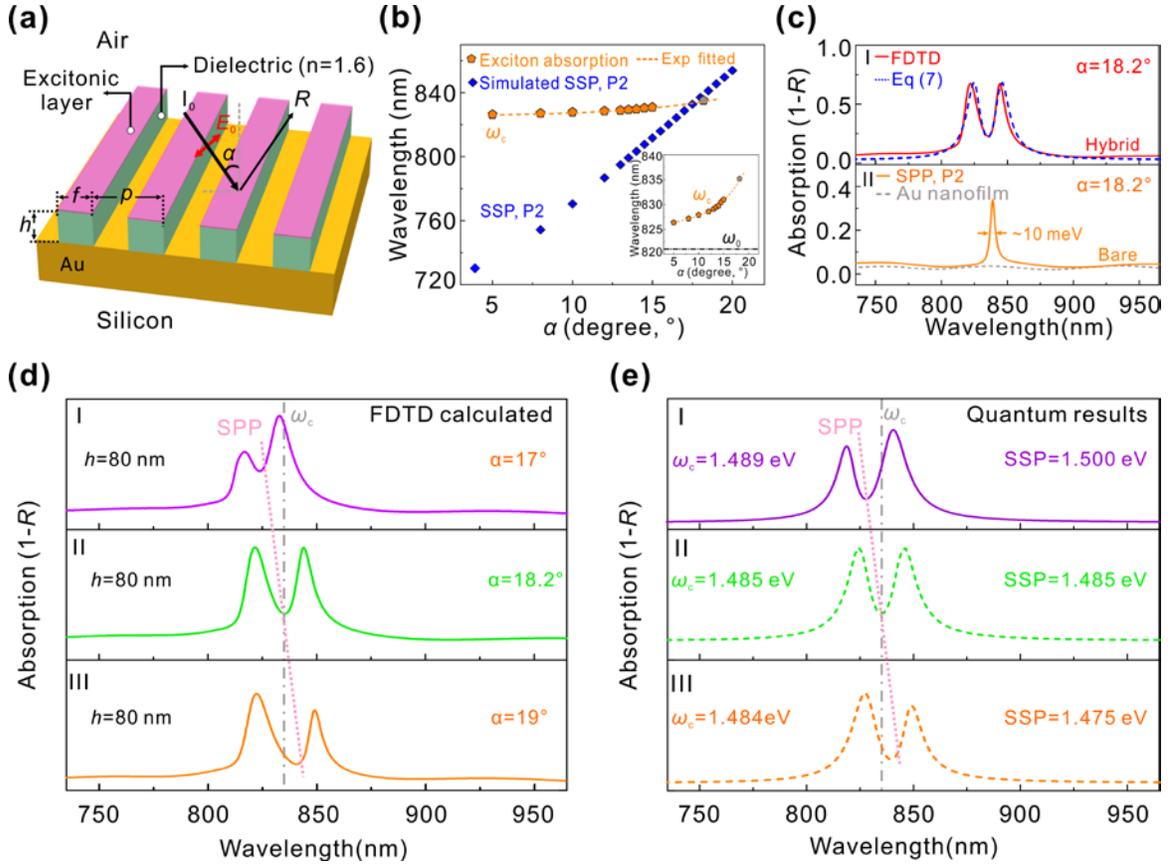

**Figure 4.** Strong coupling between the SPP mode P2 and exciton. (a) Schematic of the hybrid plasmonic nanostructure with the excitonic layer on the grating of the PGS ($p$=600 nm, $h$=80 and $f$=0.4), excited by plane waves with an incident angle of $α$. (b) The excitonic transition, $ω_c$, and the resonance wavelength of the SSP mode P2 as functions of $α$. The gray pentagon is the $ω_c$ extracted from the dip of the resonance spectral Rabi-splitting, where $ω_c$=$ω_d$. The inset is a more intuitive demonstration of the change for $ω_c$ versus $α$. (c) FDTD calculated absorption spectra of (I) the hybrid plasmonic nanostructure (solid red curve) and (II) the bare PGS at $α$=18.2°, respectively. The dashed blue curve in (I) is the absorption spectrum of the hybrid plasmonic nanostructure calculated using eq 7 with $A1$=0 and $A2$=1, which is normalized to the maximum absorption calculated at $g_{dc}$=0. (d) FDTD calculated absorption spectra for the hybrid plasmonic structure according to $α$=17, 18.2 and 19°, respectively. (e) Normalized absorption spectra

calculated using eq 7 with *A*1=0 and *A*2=1, corresponding to the parameters shown in (d). In calculations, $g_{dc}$ =18.25 meV is extracted from eq 11 with $\Delta_{qe}$=19 meV.

In summary, we have theoretically demonstrated the invalidity of the equality assumption that the spectral Rabi-splitting of the photons absorption is identical to the level splitting of the coupled plasmon-QEs systems. It is revealed that which one of both the level splitting and the spectral Rabi-splitting is easier to be reached highly correlates with the relative dissipation decays between the two subsystems. The strong coupling characterized by the spectral Rabi-splitting of the photon absorption manifests itself the relativity and diversity to the measured subsystems. It is easier to be observed from the subsystem with larger decay, and can be classified into pseudo-, dark-, middle-, and super-strong interaction regimes. We also present the identification method of the spectral Rabi-splitting signal channel and correct measurement method of *g*-factor for future experiment researches. Finally, we demonstrate by the FDTD numerical simulations that the proposed prototype of coupled plasmon-exciton system is a good candidate for possibly future experiment observations on the relativity and diversity of the strong coupling. Our work not only enriches the fundamental physics understanding of the strong coupling of light-matter interaction in nanophotonics filed, but also provides a powerful analytical tool for the experimental investigations on it.

**ASSOCIATED CONTENT**

**Supporting Information**. The Supporting Information is available free of charge on the ACS Publications website.

The theory of optical response of the components in the strong-coupling system of QE and plasmonic cavity (Section S1 in SI); FDTD calculations (Section S2 in SI); photon absorption spectra of the coupled plasmon-QE system, respectively observed from the plasmon mode and QE with $\gamma_c > \gamma_d$ (Figure S1); comparison of the LSP mode (P0) calculated based upon the PGS and that fitted from the absorption spectra of the hybrid plasmonic structure (Figure S2); comparison of the SSP mode (P2) calculated based upon the PGS and that fitted from the absorption spectra of the hybrid plasmonic structure (Figure S3) (PDF).

## AUTHOR INFORMATION

**Corresponding Author**

*E-mail: wangxueh@mail.sysu.edu.cn.

**Author Contributions**

X.H.W. proposed the idea and project outline. R.M.L. and Y.C.Y developed the full quantum theoretical model. R.M.L. performed the FDTD simulations. X.H.W. and R.M.L. co-wrote the paper, and all authors discussed the results and the manuscript. X.H.W supervised the project.

## ACKNOWLEDGMENT

We thank Drs. G. H. Xiao, G. H. Liu and G. Y. Chen for their assistance with the FDTD simulations, data processing and discussions. We also thank the national supercomputer center in Guangzhou for assisting with numerical calculations. This work was supported by the National Key R&D Programs of China (Grant No. 2016YFA0301300), the National Natural Science Foundations of China (Grant Nos. 11874438, 91750207, and 11761141015), The Key R&D Program of Guangdong Province (Grant No. 2018B030329001), the Natural Science Foundations of Guangdong (Grant Nos. 2018A030313722 and 2016A030312012), the Guangzhou Science and Technology Projects (201607020023) and the Open Fund of IPOC (BUPT) (Grant No. IPOC2018B007).

## ABBREVIATIONS

QE, quantum emitter; CQED, cavity quantum electrodynamics; PGS, plasmonic grating structure; PNS, plasmonnano-subsystem; FDTD, finite-difference time-domain; SPP, surface plasmon polariton; LSP, localized surface plasmon.

# Supporting Information

**Section S1: Optical response of the strongly coupled system consisting of the plasmonic mode and QE**

**S1.1 Photon absorption observed from the plasmonic mode in the strongly coupled plasmon-QE system**

For a quantum system consisting of a QE interacting with the plasmonic mode, if we assume that the plasmonic nanostructure supports a well-defined bosonic dipolar plasmon mode and neglect its higher-multipole modes, we can describe the plasmon mode with bosonic annihilation and creation operators. Additionally, we describe the QE as a fermionic system with only two possible states (i.e. ground and excited states). The Hamiltonian describing the quantum system can be written as [1]

$$H = H_0 + H_{int} + H_{decay}. \tag{S1}$$

The Hamiltonian describing the noninteracting evolution of the fermion and the plasmon is

$$H_0 = \varepsilon_d d^+ d + \varepsilon_c c^+ c, \tag{S2}$$

where $d$ and $c$ ($d^+$ and $c^+$) are the annihilation (creation) operators for the nanoparticle plasmon and the QE fermion of energies $\varepsilon_d = \hbar\omega_d$ and $\varepsilon_c = \hbar\omega_c$, respectively. The plasmon-fermion interaction is modeled by the Hamiltonian

$$H_{int} = -g_{dc}\left[d^+ c + c^+ d\right], \tag{S3}$$

where $g_{dc}$ is the plasmon-QE coupling energy (called as g-factor), which we take to be real. In order to describe the finite lifetimes of the plasmon and fermion, we add a term to the Hamiltonian that contains the inelastic interactions with a continuum of modes

$$H_{\text{decay}} = \int d\omega \hbar \omega f_d^+(\omega) f_d(\omega) + \int d\omega \hbar \omega f_c^+(\omega) f_c(\omega)$$
$$- \int d\omega \left[ v_d(\omega) f_d(\omega) d^+ + v_d^*(\omega) f_d^+(\omega) d \right] \quad \text{(S4)}$$
$$- \int d\omega \left[ v_c(\omega) f_c(\omega) c^+ + v_c^*(\omega) f_c^+(\omega) c \right],$$

where $f_d(\omega)$ and $f_c(\omega)$ are the annihilation operators of the continuum modes that couple to the plasmon and fermion, respectively. The corresponding coupling constants are $v_d(\omega)$ and $v_c(\omega)$.

The optical spectrum induced by the plasmons absorption can be found from Zubarev's Green function [2] $\langle\langle d; d^+ \rangle\rangle_{\omega+i0^+}$ with $\eta=1$. With the commutator relation of $dd^+ - d^+d = 1$, we get

$$\hbar(\omega+i0^+)\langle\langle d; d^+ \rangle\rangle_{\omega+i0^+}$$
$$= \langle [d(0), d^+(0)]_1 \rangle + \langle\langle [d,H]; d^+ \rangle\rangle_{\omega+i0^+}$$
$$= 1 + \langle\langle [d,H_0]; d^+ \rangle\rangle_{\omega+i0^+} + \langle\langle [d,H_{\text{int}}]; d^+ \rangle\rangle_{\omega+i0^+} + \langle\langle [d,H_{decay}]; d^+ \rangle\rangle_{\omega+i0^+} \quad \text{(S5)}$$
$$= 1 + \varepsilon_d \langle\langle d; d^+ \rangle\rangle_{\omega+i0^+} - g_{dc} \langle\langle c; d^+ \rangle\rangle_{\omega+i0^+} - \int d\omega' v_d(\omega') \langle\langle f_d(\omega'); d^+ \rangle\rangle_{\omega+i0^+}$$

Thus we have

$$\left[ \hbar(\omega+i0^+) - \varepsilon_d \right] \langle\langle d; d^+ \rangle\rangle_{\omega+i0^+}$$
$$= 1 - g_{dc} \langle\langle c; d^+ \rangle\rangle_{\omega+i0^+} - \int d\omega' v_d(\omega') \langle\langle f_d(\omega'); d^+ \rangle\rangle_{\omega+i0^+}. \quad \text{(S6)}$$

We thus need to calculate $\langle\langle c; d^+ \rangle\rangle_{\omega+i0^+}$ and $\langle\langle f_d(\omega'); d^+ \rangle\rangle_{\omega+i0^+}$. For $\langle\langle c; d^+ \rangle\rangle_{\omega+i0^+}$, we have

$$\hbar(\omega+i0^+)\langle\langle c; d^+\rangle\rangle_{\omega+i0^+}$$

$$=\langle[c(0),d^+(0)]_\eta\rangle+\langle\langle[c,H]; d^+\rangle\rangle_{\omega+i0^+}$$

$$=\langle\langle[c,H_0]; d^+\rangle\rangle_{\omega+i0^+}+\langle\langle[c,H_{\text{int}}]; d^+\rangle\rangle_{\omega+i0^+}+\langle\langle[c,H_{\text{decay}}]; d^+\rangle\rangle_{\omega+i0^+}$$

$$=\varepsilon_c\langle\langle(1-2c^+c)c; d^+\rangle\rangle_{\omega+i0^+}-g_{dc}\langle\langle(1-2c^+c)d; d^+\rangle\rangle_{\omega+i0^+} \qquad (S7)$$

$$-\int d\omega' v_c(\omega')\langle\langle(1-2c^+c)f_c(\omega'); d^+\rangle\rangle_{\omega+i0^+}$$

$$=\varepsilon_c\langle\langle c; d^+\rangle\rangle_{\omega+i0^+}-g_{dc}\langle\langle(1-2c^+c)d; d^+\rangle\rangle_{\omega+i0^+}$$

$$-\int d\omega' v_c(\omega')\langle\langle(1-2c^+c)f_c(\omega'); d^+\rangle\rangle_{\omega+i0^+}.$$

The presence of the operator $c^+c$ is the result of the fermionic character of the QE. Since the fermion has only ground and excited states, the term $\langle\langle c^+cc; d^+\rangle\rangle_{\omega+i0^+}$ vanishes, there are still new Green functions emerge that need to be calculated. The iteration of this process would produce an infinite hierarchy of equations of motion. Here we truncate it at this point by approximating the operator $c^+c$ by its expectation value $\langle c^+c\rangle=n_c$. We can thus get

$$\left[\hbar(\omega+i0^+)-\varepsilon_c\right]\langle\langle c; d^+\rangle\rangle_{\omega+i0^+}$$

$$=-g_{dc}(1-2n_c)\langle\langle d; d^+\rangle\rangle_{\omega+i0^+}-(1-2n_c)\int d\omega' v_c(\omega')\langle\langle f_c(\omega'); d^+\rangle\rangle_{\omega+i0^+}. \qquad (S8)$$

Now we need to deal with Green functions containing information related to the inelastic decay process. With the Hamiltonian we get

$$\left[\hbar(\omega+i0^+)-\hbar\omega'\right]\langle\langle f_d(\omega'); d^+\rangle\rangle_{\omega+i0^+}=-v_d^*(\omega')\langle\langle d; d^+\rangle\rangle_{\omega+i0^+}, \qquad (S9)$$

$$\left[\hbar(\omega+i0^+)-\hbar\omega'\right]\langle\langle f_c(\omega'); d^+\rangle\rangle_{\omega+i0^+}=-v_c^*(\omega')\langle\langle c; d^+\rangle\rangle_{\omega+i0^+}. \qquad (S10)$$

Substituting Eq. (S9) into Eq. (S6), we have

$$\left[\hbar(\omega+i0^+)-\varepsilon_d\right]\langle\langle d; d^+\rangle\rangle_{\omega+i0^+}$$

$$=1-g_{dc}\langle\langle c; d^+\rangle\rangle_{\omega+i0^+}-\int d\omega'\frac{1}{\hbar}\frac{|v_d(\omega')|^2}{\omega'-\omega-i0^+}\langle\langle d; d^+\rangle\rangle_{\omega+i0^+}. \qquad (S11)$$

Using the identity

$$\frac{1}{\omega' - \omega \pm i0^+} = \mathrm{P}\left(\frac{1}{\omega' - \omega}\right) \mp i\pi \delta(\omega' - \omega), \tag{S12}$$

we find

$$\begin{aligned}\int d\omega' \frac{1}{\hbar} \frac{|v_d(\omega')|^2}{\omega' - \omega - i0^+} &= \mathrm{P}\int d\omega' \frac{1}{\hbar} \frac{|v_d(\omega')|^2}{\omega' - \omega} + i\pi \frac{|v_d(\omega)|^2}{\hbar} \\ &= \delta\omega_d + i\gamma_d,\end{aligned} \tag{S13}$$

where

$$\gamma_d = \pi |v_d(\omega)|^2 / \hbar \tag{S14}$$

is the half linewidth the dipolar plasmons, and $\delta\omega_d$ represents a frequency shift. We can thus transform Eq. (S11) into

$$[\hbar\omega - \varepsilon_d + \delta\omega_d + i\gamma_d]\langle\langle d; d^+\rangle\rangle_{\omega+i0^+} = 1 - g_{dc}\langle\langle c; d^+\rangle\rangle_{\omega+i0^+}. \tag{S15}$$

Dealing in a similar way with the remaining fermionic decay channels and substituting Eq. (S10) into Eq. (S8), we can obtain

$$[\hbar\omega - \varepsilon_c + (1 - 2n_c)(\delta\omega_c + i\gamma_c)]\langle\langle c; d^+\rangle\rangle_{\omega+i0^+} = -g_{dc}(1 - 2n_c)\langle\langle d; d^+\rangle\rangle_{\omega+i0^+}. \tag{S16}$$

With Eq. (S15) and Eq. (S16), we can derive the equation of motion for $\langle\langle d; d^+\rangle\rangle_{\omega+i0^+}$ as

$$\langle\langle d; d^+\rangle\rangle_{\omega+i0^+} = \left\{\hbar\omega - \varepsilon_d + \delta\omega_d + i\gamma_d - \frac{g_{dc}^2(1 - 2n_c)}{\hbar\omega - \varepsilon_c + (1 - 2n_c)(\delta\omega_c + i\gamma_c)}\right\}^{-1}. \tag{S17}$$

We can thus obtain the photon absorption spectrum of the strongly coupled plasmon-QE system observed from the plasmonic mode as

$$\sigma_d(\omega) \propto -\mathrm{Im}\left\{\hbar\omega - \varepsilon_d + \delta\omega_d + i\gamma_d - \frac{g_{dc}^2(1 - 2n_c)}{\hbar\omega - \varepsilon_c + (1 - 2n_c)(\delta\omega_c + i\gamma_c)}\right\}^{-1}. \tag{S18}$$

We note that

$$\varepsilon_d = \hbar\omega_d,$$
$$\gamma_c = \hbar\Gamma_0/2, \tag{S19}$$
$$\gamma_d = \hbar\kappa/2,$$

the right terms are the familiar ones with the rad/s unit; $\kappa$ is the decay rate of the plasmon mode.

The spontaneous emission rate of the QE in vacuum is [S3]

$$\Gamma_0 = \frac{\omega_0^3 d_c^2}{3\pi\hbar\varepsilon_0 c^3}. \tag{S20}$$

If we take the system to be initially prepared in the ground state, then we have $n_c = 0$ for the QE.

We can thus calculate the photon absorption spectrum as

$$\sigma_d(\omega) \propto -\mathrm{Im}\frac{1}{\hbar\omega - \varepsilon_d + i\gamma_d - \dfrac{g_{dc}^2}{\hbar\omega - \varepsilon_c + i\gamma_c}}. \tag{S21}$$

For the situation that the frequency of the QEs is on resonance with the frequency of the plasmon mode, i.e., $\varepsilon_c = \varepsilon_d$, we have

$$\sigma_d(\omega) \propto \frac{(\hbar\omega - \varepsilon_d)^2 \gamma_d + \left[g_{dc}^2 + \gamma_d\gamma_c\right]\gamma_c}{\left[(\hbar\omega - \varepsilon_d)^2 - g_{dc}^2 - \gamma_d\gamma_c\right]^2 + (\hbar\omega - \varepsilon_d)^2 (\gamma_c + \gamma_d)^2}, \tag{S22}$$

and therefore

$$\frac{\mathrm{d}\sigma_d(\omega)}{\mathrm{d}\omega} \propto -\hbar\frac{2\gamma_d(\hbar\omega - \varepsilon_d)\left[(\hbar\omega - \varepsilon_d)^4 + s_1(\hbar\omega - \varepsilon_d)^2 + s_2\right]}{\left\{\left[(\hbar\omega - \varepsilon_d)^2 - (g_{dc}^2 + \gamma_c\gamma_d)\right]^2 + (\hbar\omega - \varepsilon_d)^2 \gamma_c\gamma_d\right\}^2}, \tag{S23}$$

where

$$s_1 = 2\frac{\gamma_c}{\gamma_d}\left(g_{dc}^2 + \gamma_c\gamma_d\right), \quad s_2 = \frac{1}{\gamma_d}\left[\gamma_c^3 - (2\gamma_c + \gamma_d)g_{dc}^2\right]\left(g_{dc}^2 + \gamma_c\gamma_d\right). \tag{S24}$$

Let $\dfrac{\mathrm{d}\sigma_d(\omega)}{\mathrm{d}\omega} = 0$, we get $\hbar\omega = \varepsilon_c$, or

$$(\hbar\omega - \varepsilon_c)^2 = \frac{-s_1 \pm \sqrt{s_1^2 - 4s_2}}{2} \tag{S25}$$

In order to satisfy Eq. (S25), we should have $s_2 < 0$, we thus get the condition for observing the mode splitting in the absorption spectrum observed from the plasmonic mode is

$$g_{dc}^2 > g_{pl}^2 = \frac{\gamma_c^2}{2(1+\gamma_d/2\gamma_c)}. \tag{S26}$$

For this condition, the energies of the two mixed levels observed in the absorption spectrum can be derived as

$$\varepsilon_\pm = \varepsilon_c \pm \Delta_{pl}, \quad \Delta_{pl} = \sqrt{g_{dc}(1+\gamma_c/\gamma_d)\cdot(g_{dc}^2+\gamma_c\gamma_d)^{1/2} - \gamma_c/\gamma_d(g_{dc}^2+\gamma_c\gamma_d)} \tag{S27}$$

**S1.2 Photon absorption observed from the QE in the strongly coupled plasmon-QE system**

Now we consider the Green function $\langle\langle c; c^+ \rangle\rangle_{\omega+i0^+}$ with $\eta = -1$. With the commutator relation $cc^+ + c^+c = 1$, we get

$$\begin{aligned}
&\hbar(\omega+i0^+)\langle\langle c; c^+\rangle\rangle_{\omega+i0^+} \\
&= \langle[c(0), c^+(0)]_{-1}\rangle + \langle\langle[c,H]; c^+\rangle\rangle_{\omega+i0^+} \\
&= 1 + \langle\langle[c,H_0]; c^+\rangle\rangle_{\omega+i0^+} + \langle\langle[c,H_{int}]; c^+\rangle\rangle_{\omega+i0^+} + \langle\langle[c,H_{decay}]; c^+\rangle\rangle_{\omega+i0^+} \\
&= 1 + \varepsilon_c \langle\langle(1-2c^+c)c; c^+\rangle\rangle_{\omega+i0^+} - g_{dc}\langle\langle(1-2c^+c)d; c^+\rangle\rangle_{\omega+i0^+} \\
&\quad - \int d\omega' v_c(\omega')\langle\langle(1-2c^+c)f_c(\omega'); c^+\rangle\rangle_{\omega+i0^+} \\
&= 1 + \varepsilon_c \langle\langle c; c^+\rangle\rangle_{\omega+i0^+} - g_{dc}\langle\langle(1-2c^+c)d; c^+\rangle\rangle_{\omega+i0^+} \\
&\quad - \int d\omega' v_c(\omega')\langle\langle(1-2c^+c)f_c(\omega'); c^+\rangle\rangle_{\omega+i0^+}
\end{aligned} \tag{S28}$$

Here we also approximate the operator $c^+c$ by its expectation value $\langle c^+c\rangle = n_c$, we can thus get

$$\begin{aligned}
&[\hbar(\omega+i0^+)-\varepsilon_c]\langle\langle c; c^+\rangle\rangle_{\omega+i0^+} \\
&= 1 - g_{dc}(1-2n_c)\langle\langle d; c^+\rangle\rangle_{\omega+i0^+} - (1-2n_c)\int d\omega' v_c(\omega')\langle\langle f_c(\omega'); c^+\rangle\rangle_{\omega+i0^+}.
\end{aligned} \tag{S29}$$

We need to calculate $\langle\langle d; c^+\rangle\rangle_{\omega+i0^+}$ and $\langle\langle f_d(\omega'); c^+\rangle\rangle_{\omega+i0^+}$. For $\langle\langle d; c^+\rangle\rangle_{\omega+i0^+}$, then we have

$$\hbar(\omega+i0^+)\langle\langle d;c^+\rangle\rangle_{\omega+i0^+}$$
$$=\langle[d(0),c^+(0)]_{-1}\rangle+\langle\langle[d,H];c^+\rangle\rangle_{\omega+i0^+}$$
$$=2\rho_{dc}+\langle\langle[d,H_0];c^+\rangle\rangle_{\omega+i0^+}+\langle\langle[d,H_{int}];c^+\rangle\rangle_{\omega+i0^+}+\langle\langle[d,H_{decay}];c^+\rangle\rangle_{\omega+i0^+} \quad (S30)$$
$$=\varepsilon_d\langle\langle d;c^+\rangle\rangle_{\omega+i0^+}-g_{dc}\langle\langle c;c^+\rangle\rangle_{\omega+i0^+}-\int d\omega' v_d(\omega')\langle\langle f_d(\omega');c^+\rangle\rangle_{\omega+i0^+}$$

Thus we have

$$\left[\hbar(\omega+i0^+)-\varepsilon_d\right]\langle\langle d;c^+\rangle\rangle_{\omega+i0^+}$$
$$=-g_{dc}\langle\langle c;c^+\rangle\rangle_{\omega+i0^+}-\int d\omega' v_d(\omega')\langle\langle f_d(\omega');c^+\rangle\rangle_{\omega+i0^+}. \quad (S31)$$

We can also get

$$\left[\hbar(\omega+i0^+)-\hbar\omega'\right]\langle\langle f_d(\omega');c^+\rangle\rangle_{\omega+i0^+}=-v_d^*(\omega')\langle\langle d;c^+\rangle\rangle_{\omega+i0^+}, \quad (S32)$$
$$\left[\hbar(\omega+i0^+)-\hbar\omega'\right]\langle\langle f_c(\omega');c^+\rangle\rangle_{\omega+i0^+}=-v_c^*(\omega')\langle\langle c;c^+\rangle\rangle_{\omega+i0^+}. \quad (S33)$$

Substituting Eq. (S33) into Eq. (S29), we have

$$\left[\hbar(\omega+i0^+)-\varepsilon_c\right]\langle\langle c;c^+\rangle\rangle_{\omega+i0^+}$$
$$=1-g_{dc}(1-2n_c)\langle\langle d;c^+\rangle\rangle_{\omega+i0^+}-(1-2n_c)\int d\omega'\frac{1}{\hbar}\frac{|v_c(\omega')|^2}{\omega'-\omega-i0^+}\langle\langle c;c^+\rangle\rangle_{\omega+i0^+}. \quad (S34)$$

We thus have

$$\left[\hbar\omega-\varepsilon_c+(1-2n_c)(\delta\omega_c+i\gamma_c)\right]\langle\langle c;c^+\rangle\rangle_{\omega+i0^+}=1-g_{dc}(1-2n_c)\langle\langle d;c^+\rangle\rangle_{\omega+i0^+}. \quad (S35)$$

Substituting Eq. (S32) into Eq. (S31), we have

$$\left[\hbar(\omega+i0^+)-\varepsilon_d\right]\langle\langle d;c^+\rangle\rangle_{\omega+i0^+}$$
$$=-g_{dc}\langle\langle c;c^+\rangle\rangle_{\omega+i0^+}-\int d\omega'\frac{1}{\hbar}\frac{|v_d(\omega')|^2}{\omega'-\omega-i0^+}\langle\langle d;c^+\rangle\rangle_{\omega+i0^+}. \quad (S36)$$

We thus have

$$\left[\hbar\omega-\varepsilon_d+\delta\omega_d+i\gamma_d\right]\langle\langle d;c^+\rangle\rangle_{\omega+i0^+}=-g_{dc}\langle\langle c;c^+\rangle\rangle_{\omega+i0^+}. \quad (S37)$$

With Eq. (S35) and Eq. (S37), we can obtain the equation of motion for $\langle\langle c;c^+\rangle\rangle_{\omega+i0^+}$ as

$$\langle\langle c; c^+\rangle\rangle_{\omega+i0^+} = \left\{\hbar\omega - \varepsilon_c + (1-2n_c)(\delta\omega_c + i\gamma_c) - \frac{g_{dc}^2(1-2n_c)}{\hbar\omega - \varepsilon_d + \delta\omega_d + i\gamma_d}\right\}^{-1}. \qquad (S38)$$

If we consider the optical absorption spectrum obtained from $\langle\langle c; c^+\rangle\rangle_{\omega+i0^+}$, we have the photon absorption spectrum of the strongly coupled plasmon-QE system observed from the QE as

$$\sigma_c(\omega) \propto -\text{Im}\left\{\langle\langle c; c^+\rangle\rangle_{\omega+i0^+}\right\}$$
$$= -\text{Im}\left\{\hbar\omega - \varepsilon_c + (1-2n_c)(\delta\omega_c + i\gamma_c) - \frac{g_{dc}^2(1-2n_c)}{\hbar\omega - \varepsilon_d + \delta\omega_d + i\gamma_d}\right\}^{-1}. \qquad (S39)$$

If we take the system to be initially prepared in the ground state, then we have $n_c = 0$ for the QE. We also assume that $\delta\omega_d$ and $\delta\omega_c$ have already been accounted in $\varepsilon_d$ and $\varepsilon_c$ by renormalization, respectively. We thus have

$$\sigma_c(\omega) \propto -\text{Im}\frac{1}{\hbar\omega - \varepsilon_c + i\gamma_c - \dfrac{g_{dc}^2}{\hbar\omega - \varepsilon_d + i\gamma_d}} \qquad (S40)$$

If without the plasmons, then $g_{dc} = 0$, we thus get the familiar Lorentz line shape of the QE

$$\sigma_c(\omega) \propto -\text{Im}\left\{\omega - \omega_c + i\frac{\Gamma_0}{2}\right\}^{-1} = \frac{\Gamma_0/2}{(\omega - \omega_c)^2 + (\Gamma_0/2)^2}. \qquad (S41)$$

If $g_{dc} \neq 0$, for the situation that the frequency of the QEs is resonant with the frequency of the plasmon mode, i.e., $\varepsilon_c = \varepsilon_d$, we have

$$\sigma_c(\omega) \propto \frac{(\hbar\omega - \varepsilon_c)^2 \gamma_c + \left[g_{dc}^2 + \gamma_c\gamma_d\right]\gamma_d}{\left[(\hbar\omega - \varepsilon_c)^2 - g_{dc}^2 - \gamma_c\gamma_d\right]^2 + (\hbar\omega - \varepsilon_c)^2(\gamma_c + \gamma_d)^2}, \qquad (S42)$$

and therefore,

$$\frac{d\sigma_c(\omega)}{d\omega} \propto -\hbar\frac{2\gamma_c(\hbar\omega - \varepsilon_c)\left[(\hbar\omega - \varepsilon_c)^4 + s_1'(\hbar\omega - \varepsilon_c)^2 + s_2'\right]}{\left\{\left[(\hbar\omega - \varepsilon_c)^2 - (g_{dc}^2 + \gamma_c\gamma_d)\right]^2 + (\hbar\omega - \varepsilon_c)^2(\gamma_c + \gamma_d)^2\right\}^2}, \qquad (S43)$$

where

$$s'_1 = 2\frac{\gamma_d}{\gamma_c}\left(g_{dc}^2 + \gamma_c\gamma_d\right), \; s'_2 = \frac{1}{\gamma_c}\left[\gamma_d^3 - (2\gamma_d + \gamma_c)g_{dc}^2\right]\left(g_{dc}^2 + \gamma_c\gamma_d\right). \quad (S44)$$

Let $\dfrac{d\sigma_c(\omega)}{d\omega} = 0$ we get $\hbar\omega = \varepsilon_d$, or

$$(\hbar\omega - \varepsilon_d)^2 = \frac{-s'_1 \pm \sqrt{s'^2_1 - 4s'_2}}{2} \quad (S45)$$

In order to satisfy Eq. (S45), we should have $s'_2 < 0$, we thus get the condition for observing the mode splitting in the absorption spectrum observed from the QE channel is

$$g_{dc}^2 > g_{qe}^2 = \frac{\gamma_d^2}{2(1 + \gamma_c/2\gamma_d)}, \quad (S46)$$

and the energies of the two mixed levels in absorption spectrum can be written as

$$\varepsilon_\pm = \varepsilon_d \pm \Delta_{qe}, \Delta_{qe} = \sqrt{g_{dc}(1 + \gamma_d/\gamma_c)\cdot(g_{dc}^2 + \gamma_c\gamma_d)^{1/2} - \gamma_d/\gamma_c(g_{dc}^2 + \gamma_c\gamma_d)}. \quad (S47)$$

For quantum coupling systems with dipolar plasmons and QE, we usually have $\gamma_d > \gamma_c$. Therefore, the results above indicate that observing Rabi splitting in the optical absorption spectrum from the dipolar plasmons is much easier than from the QE. On the contrary, if we have $\gamma_d < \gamma_c$, i.e., the dipolar plasmons has a lower dissipation than that of the QE, we can observe Rabi splitting more easily in the optical absorption spectrum from the QE, as shown in Fig. S1.

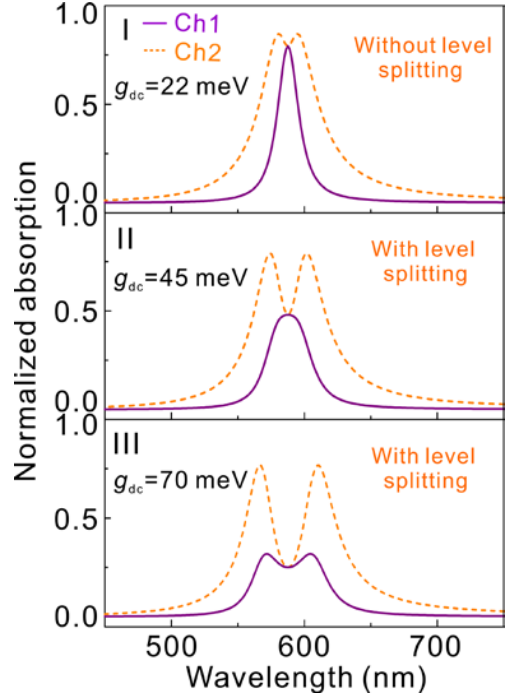

**Figure S1.** Normalized absorption of the coupled plasmon-QE system calculated using equation (7), respectively observed from the plasmon mode (Ch1, *i.e.* $A_1=1$, $A_2=0$) and the QE channel (Ch2, *i.e.* $A_1=0$, $A_2=1$), under different $g_{dc}$ values of (I) 22, (II) 45, and (III) 70 meV. In calculations, $\omega_c = \omega_d = 2.11$ eV, $\gamma_d = 25$ meV and $\gamma_c = 75$ meV, typical values for the molecule or quantum dot exciton transition at room temperature [4-6].

**Section S2: Finite-difference time-domain (FDTD) calculations**

In FDTD (Lumerical Solutions) calculations of the EF distributions and absorption spectra (1-*R*, *R*: reflection) as shown in Fig. 3, the PGS was constructed by placing a dielectric ($n=1.6$) grating with a period of $p = 600$ nm, thickness $h$ (can be changed), and fill factor $f = 0.4$ over a gold nanofilm (100 nm) which is fixed on a bulk silica substrate. Note that, in the calculation of the absorption spectrum for the bare PGS, a dielectric layer of 5 nm with the refractive index of 1.612 (corresponding to $\varepsilon_{ex}(\infty) = 2.6$ in Eq (S48)) was used to replace the excitonic thin-layer in the hybrid nanostructure. The bulk dielectric function tabulated by Johnson and Christy was used for Au. On the other hand, we used the FDTD method to calculate the optical properties of the hybrid plasmonic nanostructure with excitonic thin-layer. The excitonic thin-layer was modeled as a dispersive medium, and the dielectric permittivity of the exciton layer has been described by the Lorentz model as

$$\varepsilon_{ex}(\omega) = \varepsilon_{ex}(\infty) + \frac{f_0 \omega_0^2}{(\omega_0^2 - \omega^2 - 2i\gamma_c \omega)}, \qquad (S48)$$

where $\varepsilon_{ex}(\infty) = 2.6$ is the high-frequency component of the dielectric function, $f_0 = 0.09$ is the reduced oscillator strength, $\omega_0 = 1.642$ and $1.51$ eV are the exciton transition energies used in Figs. 3 and 4, respectively, and $2\gamma_c = 35$ meV is the linewidth of exciton absorption.

In order to show the absorption of the excitonic thin-layer without coupling to the LSP mode P0, as well as the evolution of the interaction between the two subsystems as shown in Fig. 3, we calculated the absorption spectra of the hybrid plasmonic nanostructure with varying $h$ from 60 to 200 nm, under a normal incidence ($\alpha = 0°$) (see the following Fig. S2). Similarly, to show the absorption of the excitonic layer without coupling to the SPP mode P2, as well as the evolution of the interaction between the two subsystems as shown in Fig. 4, we calculated the absorption spectra of the hybrid plasmonic nanostructure with varying $\alpha$ from $10°$ to $19°$ for TM polarized light at $h=80$ nm (see the following Fig. S3).

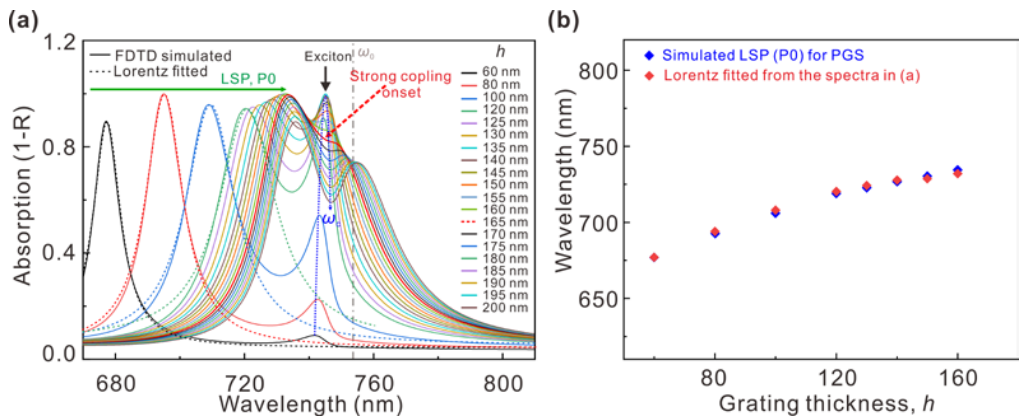

**Figure S2. Comparison of the LSP mode (P0) calculated based upon the bare PGS and that fitted from the absorption spectra of the hybrid plasmonic structure**. (**a**) Absorption spectra of the hybrid plasmonic nanostructure calculated using FDTD solutions with a varying grating thickness, $h$, from 60 to 200 nm, under the TM polarized incident light at $\alpha=0°$. (**b**) The resonance wavelength of the LSP mode P0 as a function of the grating thickness, $h$. Blue rhombuses: results are directly simulated using FDTD based on the bare PGS. Red rhombuses: results are extracted from the spectra depicted in (**a**) by Lorentz fitting.

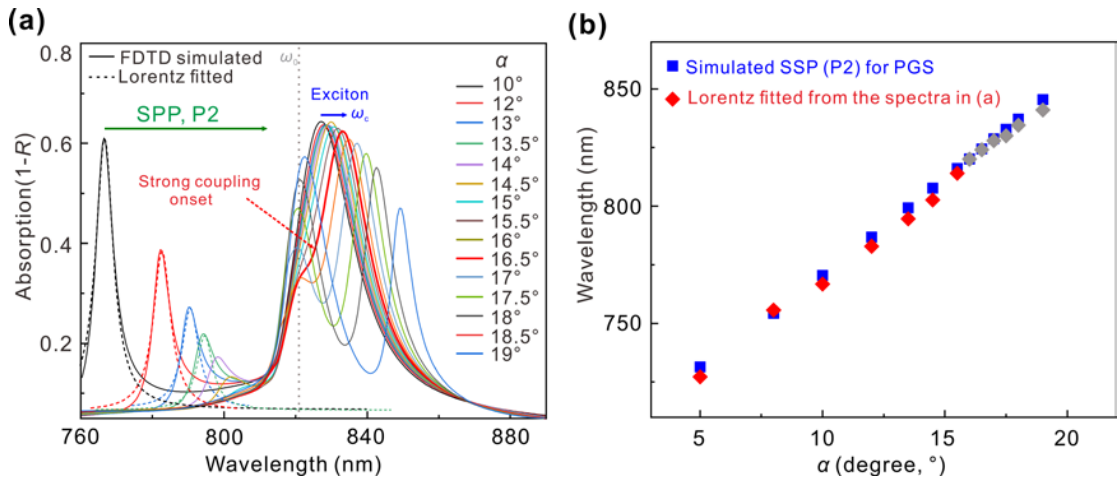

**Figure S3. Comparison of the SSP mode (P2) calculated based upon the bare PGS and that fitted from the absorption spectra of the hybrid plasmonic structure**. (**a**) Absorption spectra of the hybrid plasmonic nanostructure calculated using FDTD solutions with a varying $\alpha$ from 10 to 19°, under the TM polarized incident light at $h = 80$ nm. (**b**) The resonance wavelength of the SSP mode (P2) as a function of $\alpha$. The blue rhombuses: results are directly simulated based on the bare PGS at different $\alpha$. Red rhombuses: results are extracted from the spectra in (**a**) by Lorentz fitting. The gray rhombuses are results extracted from the dips of the spectral Rabi-splitting in the absorption spectra shown in (**a**).